\documentclass[runninghead]{llncs}
\usepackage{graphicx}
\usepackage{array}
\usepackage{xspace}
\usepackage{graphicx}
\usepackage{url}
\usepackage {mathtools}
\usepackage[table]{xcolor}
\usepackage{float}
\usepackage{subfigure}
\usepackage{booktabs}
\usepackage{indentfirst} 
\usepackage{comment}
\usepackage{color}

\def\ie{\textit{i.e.}\xspace}

\def\etc{\textit{etc.}\xspace}
\def\eg{\textit{e.g.}\xspace}

\begin{document}
\title{Long-Range Gesture Recognition Using Millimeter Wave Radar}

\author{
Yu Liu\inst{1} \orcidID{0000-0002-2129-1789}\and
Yuheng Wang\inst{1}\orcidID{0000-0001-6969-9612} \and
Haipeng Liu\inst{1} \orcidID{0000-0002-8045-6416}\and
Anfu Zhou\inst{1}\orcidID{0000-0002-8785-3350}\and
Jianhua Liu\inst{2}\orcidID{0000-0003-0107-686X}\and
Ning Yang\inst{2}\orcidID{0000-0003-1396-3423}
}
\authorrunning{Y. Liu et al.}
\institute{Beijing University of Posts and Telecommunications, China
\and
OPPO Co. Ltd., Beijing, China 
}

\maketitle

\begin{abstract}
Millimeter wave (mmWave) based gesture recognition technology
provides a good human computer interaction (HCI) experience. 
Prior works focus on the close-range gesture recognition, but fall short in range extension, \ie, they are unable to recognize gestures more than one meter away from considerable noise motions.
In this paper, we design a long-range gesture recognition model which utilizes a novel data processing method and a customized artificial 
Convolutional Neural Network (CNN).
Firstly, we break down gestures into multiple reflection points and extract their spatial-temporal features
which depict gesture details.
Secondly, we design a CNN to learn changing patterns of extracted features respectively and 
output 
the recognition
result. 
We thoroughly evaluate our proposed system by implementing on a commodity 
mmWave radar.
Besides, we also provide more extensive assessments to demonstrate that the proposed system is practical in several real-world scenarios. 

\end{abstract} 

\section{Introduction}
\setlength{\parindent}{2em} 
Contactless gesture recognition is a popular approach to realize natural human-computer interaction (HCI) for a better experience, so
more and more external physical gesture devices \cite{18,19} will be replaced by ``in air" gestures. 
To realize this HCI, researchers focus on wireless signal sensing.
Compared with candidate sensing methods (\eg, WiFi signal \cite{9}, sonic wave \cite{11}, and ultrusonic wave \cite{12}), millimeter wave (mmWave) is sensitive to detect tiny variations, \ie, centimiter-level finger movements.
While visible \cite{6} and infrared {\cite{7} light sensing are much more accurate on imaging a hand and thus gesture recognition, mmWave has unique advantages of privacy protection and energy consumption.

Therefore, mmWave is the most suitable choice for contactless gesture recognition.
Especially, 
mmWave 
is mainly used in 5G technology \cite{1}, so
it will not only be a new radio access standard but a potential sensing tool. 
The research on mmWave gesture recognition has obtained many achievements. 
At the practical application level, it can be used in the automotive industry to provide a safe and intuitive control interface for drivers. However, 
Not all gestures of passengers sitting in the car can be accurately recognized, and the model will be low accuracy due to the interference of distance and the items in the car\cite{2}.
In addition, a short-range compact 60GHz mmWave radar sensor that is sensitive to fine dynamic hand motions has been created. But this model can only be used in a limited diatance\cite{3}.
Therefore, prior work using mmWave radar is limited by the distance problem, which is an unavoidable challenge for us.
Specifically, they can only realize gesture recognition under a short range situation, if the range is extended, the radar receives more reflective signal information, and the traditional method can not separate the interferer information and the effective information, thus distant gestures can not be recognized accurately.

In this paper, we design a long-range gesture recognition model with a customized Convolutional Neural Network(CNN).
Firstly, in this model, we realize the accurate recognition of long-range gestures. Based on the basic principle of radar, we analyze the expressions of transmitting signals and reflected signals. We can accurately judge the distance, angle, velocity and other information of hands through the calculation of Fast Fourier Transformation(FFT) and Range-Doppler(R-D) algorithm. As long as the object is within the detection range of radar signal, we can realize the accurate location of the object. 
Secondly, our CNN is able to classify gestures correctly. In the previous CNN, all data is often processed in the same layer, which leads to some interference data to affect the result. However, we divide our CNN into five sub-layers, each layer processes the corresponding features, with less interference, and finally combine the results for judgment.
Different layers of CNN handle different features separately, so they will not interfere with each other. It ensures that the model will not be disturbed by the invalid information when processing the features. The model can maintain high accuracy even if the extended distance leads to more interference.

We evaluate the proposed model thoroughly by implementing on a 
mmWave radar. 
Our experiments demonstrate that the accuracy of the model is sufficient 
to 
be applied in certain real-world situations. 
Moreover, we conduct various extensive tests to explore the influence of different factors on the accuracy of the model.
Specifically, we simulate a family living room by placing appropriate amounts of furnitures (\eg,
televisions, chairs) to make the real domestic facilities.
Then we arrange participants to perform four gestures which are designed in advance and get the accuracy of the model judgement to evaluate the practicability of our model.

In conclusion, our contributions are as follows:

\emph{(i)} We construct 
a long-range gesture recognition model 
by extracting the spatial-temporal features of hands' reflection points.
Then we design a CNN to learn the points' features for recognition. 

\emph{(ii)} We utilize a 
mmWave radar sensor to implement the proposed model
and thus recognize gestures automatically. 

\emph{(iii)} We verify that our proposed model is robust within several real-world situations ({\eg, family living room, multiple people, and real-time situation}) and discuss the current limitations with several potential solutions.

\section{Related Work}
\setlength{\parindent}{2em} 
\textbf{mmWave sensing:} 
Nowadays, mmWave has widely been used for 
recognition
({\eg, vehicular communications\cite{5}, drone tracking\cite{14}, material identification\cite{15}}). In addition, because of its large bandwidth, it can sense small changes in motion.
This feature is often used in applications such as finger tracking\cite{16}: combining effcient, dynamic path tracking algorithms and radar to track pathing; Gesture Recognition\cite{4}: utilizing the feasibility of human gesture recognition using the spectra of radar measurement parameters for recognition. But the disadvantage of these methods is ignoring the distance between users and radar. Besides, the calculation of algorithms is too complicated and slow.

\textbf{Other long-range gesture recognition solutions:}

\emph{(i) Visible Light:}
Using visible light for communication and perception is a low-cost, green and low-carbon technology.
At the same time, it also has the advantage of avoiding the interference of other electromagnetic waves.
For example, LiGest\cite{6}, is a hand gesture recognition model based on visible light. The general idea is that different gestures' shadows move in unique patterns. Most models based on visible light have a high accuracy rate, but their application is unlikely due to the low penetration of visible light and the high cost of optical instruments.

\emph{(ii) Infrared Light:}
Infrared light is not in direct contact with the object under test, and has the advantage of high sensitivity and quick response.
At present, there are two popular infrared sensor electronic devices: Leapmotion and Kinect.
Leapmotion captures a hand motion in 3D, analyzes it, and carries out motion control on the contents using its interenal infrared camera\cite{7}. However, Leapmotion is not able to determine all attributes for each frame.
Kinect is another model to get depth images. The advantage is that it can detect up to six people, including recognizing simultaneously motions of two people\cite{8}.

\emph{(iii) WiFi signal:}
WiFi signal is widely used in our daily life, 
including mobile device networking, wireless sensor networks\cite{19,20} and also gesture recognition.
Many models utilized WiFi signal 
has been applied for daily life.
The latest study proposes WiMU, a WiFi based multi-users gesture recognition system\cite{9}. It solves the problem that previous WiFi models can not be used with multiple people, and provides a lower accuracy. Comparing to mmWave, the system is not sensitive enough to recognize complex gestures.

\emph{(iv) Sonic and Ultrasonic Wave:}
Some studies have implemented gesture recognition using sound waves. The main idea is to combine the Doppler effect and the division of power levels short-time Fourier transforms on the frequency domain to recognize gestures\cite{10}. Moreover, some studies use a linearly frequency modulated ultrasonic signal. They estimate range and receive signal strength (RSS) of reflected signal to recognize gestures\cite{11}. 
But the system is not stable enough, with accuracies in the range 64.5-96.9\%, based on the number of gestures to be recognized\cite{12}. Besides, the bandwidth characteristic of ultrasonic wave leads to a low sensitivity. However the results are affected by the external environment factors, both noise and interference gestures have a bad influence on the accuracy of the method.

\section{Overview}
\setlength{\parindent}{2em} 
The proposed model aims at an attempt of mmWave based gesture recognition in a long-range scenario, which can be applied to many aspects in our daily life. 
Specifically, the model can cooperate with smart home system to provide better user experience, \eg, controling smart home appliances by ``in air" gesture at a distance. 
For instance, 
a user can directly wave her hands up and down to control the light’s state instead of touching the switch. 
Moreover, when a user is sitting on the couch and wants to draw curtains in the distance, what she only has to do is pointing at the radar and swiping her hand left or right. 
Therefore, mmWave based long-range gesture recognition
is able to replace multiple 
unnecessary and troublesome movements, 
and acquire more comfortable and convenient experience.

Our model consists of following three modules which are shown in Fig.\ref{overview}: 

\emph{(i) Signal Transforamtion:} 
This module captures reflected mmWave signals on gesturing hand and feeds them into the next \emph{Information Extraction} module.
In this module, we use the 
mmWave radar sensor to send the FMCW signal. When the signal arrives at hands, it will be reflected and received by the radar reciever. Then, the signal will be input into the next module after certian pre-processing methods. 
 
\emph{(ii) Information Extraction:} This module constructs the signal into a gesture point cloud model and provides it to the next \emph{Neural Network} module for recognition.
In the point cloud model, each of the inner point has its own five features,
namely the x-y-z coordinate of the reflection point, velocity, and intensity. From the point cloud, we can also observe the trend of gestures' change clearly.

\emph{(iii) Convolutional Neural Network:} This module learns the changing pattern of the point cloud model and returns a classification result of the gesture.
We input the point cloud data into this module. In advance, we creates an customized artificial CNN for handling the five features, every feature has its own layers. After the data passing CNN, the system will give the type of the gestures.

\begin{figure}
	\includegraphics[width=\textwidth]{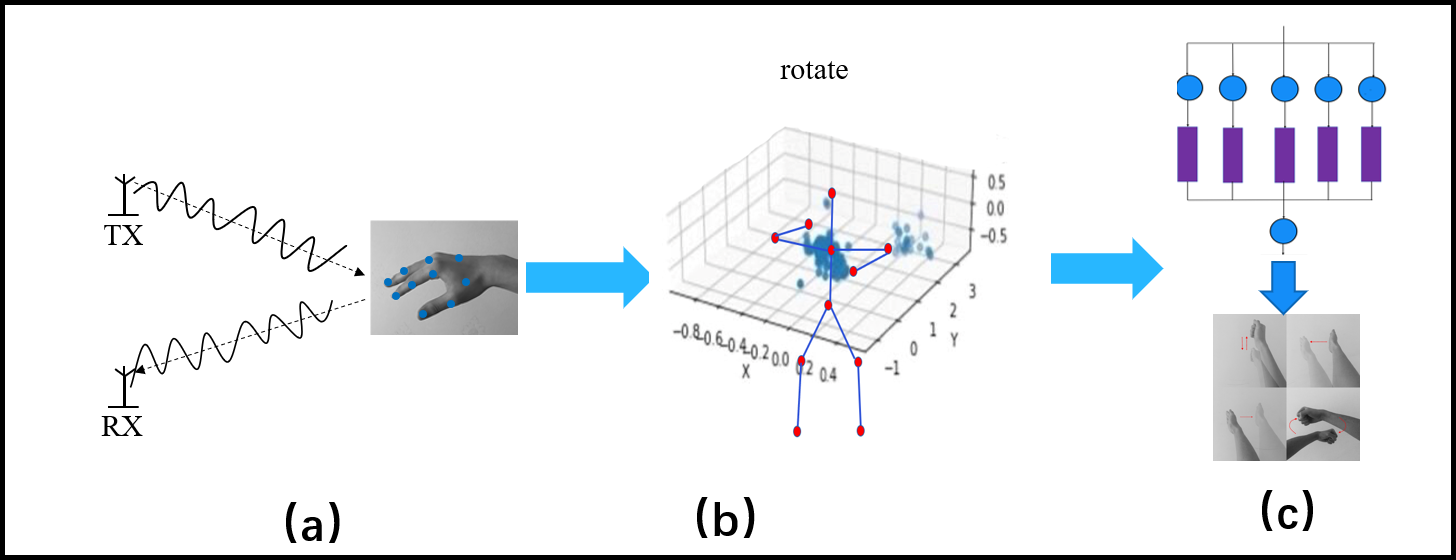}
	\caption{The model consists of three modules: (a) Signal Transformation (b)Information Extraction (c)Neural Network} \label{overview}
\end{figure}
In this work, our \textbf{main contributions} lie in last two modules. 
Through the \emph{Information Extraction} module, 
our model is able to construct gesture trends in space, so as to visually observe gesture types and help us to classify them later. In previous works, however, those models usually use raw data without being processed, resulting in a low accuracy.
The second one lies in the \emph{Convolutional Neural Network} module, by taking advantage of multi-branch integration network architecture. We can distinguish multi-category gesture information with high accuracy. 
This approach achieves the high accuracy and classification effect that the normal network used in the past work cannot be obtained.
\section{Methodology}

As noted before, our method consists of gesture point cloud model and the customized CNN.
\emph{Firstly, }the model can depict
the changing trend of gestures in space, 
 however, there is a challenge existing when constructing the model:
Long-range gestures will be interfered by some useless signals occurred by any surrounding object which reflects the mmWave signals, because the mmWave radar is omnidirectional. The problem will adversely impact the feature extraction process.
Specifically, previous studies set the distance between the radar and the user 
very small, 
so there are less interfering objects in the detection range of radar and we will lose much effetive information at the meanwhile.
\emph{Secondly,} leaveraging on the model, we custom-design a CNN to classify gestures automatically.
The design of CNN is one of the most innovative and challenging part of our solution:
The CNN is responsible to process all features comprehensively and analyze the most reasonable changing trend, but we find that common convolutional neural networks cannot be used.
Specifically, traditional convolutional neural network, such as VGG \cite{5}, have high network complexity and require too many parameters. 
Besides, it can be proved by our preliminary experiments that as the network deepens, vanishing gradient problem will appear, and lead to network degradation problems accordingly.
To meet the challenge, we propose two following methods: Firstly, 
the point cloud model we built contains valid information points and useless information points generated by the reflection of other interfering objects. These useless points will interfere with our later calculation, thus affecting the accuracy of the model, so we adopt CA-CFAR technique.
With this technique, dynamic hot points with valid information can be selected after the Range-Doppler (R-D) image is calculated, so we obtain features such as the distance, velocity and energy intensity of each spot. Thus these signal information is extracted from interference effectively.}

Secondly, we recommend a novel 
multi-branch CNN 
architecture instead of common networks used in the past. 
It not only solves the problems caused by traditional networks, but also gets better learning by passing each feature through a layer. 
We analyze features by layering to avoid cross interference between data, so as to obtain better accuracy.
\subsection{Primer: Radar Fundamental}

The proposed system adopts the radar principle that we percieve the gestures by analyzing the difference between the transmitted and the reflection signals from the hand.
Firstly, we suppose the expressions of the transmitted signal Equ.\ref{trans} and the reflected signal Equ.\ref{receive}  are as follows:
\begin{equation}
x_{1}=sin(\alpha_{1}\times t+\phi_{1}),\label{trans}
\end{equation}
\begin{equation}
x_{2}=sin(\alpha_{2}\times t+\phi_{2}),\label{receive}
\end{equation}

To obtain the distance information, we use a mixer, a component of the radar system, to combine transmitted and reflected signals to generate an intermediate frequency signal (IF) Equ.\ref{IF}.
In the 
, d is the distance between object and radar; c is the lighting speed.
\begin{equation}
x_{if}=Asin(2\pi f_{0}t+\phi_{0}), \label{IF}
 f_{0}=\frac{S2d}{c} 
, \phi_{0}=\frac{4\pi d}{\lambda}
\end{equation}
After we get the expression of the IF signal, we can compute the distance of the radar to the target object.
When we take the gesture as the target object, the velocity and intensity value of each point are obtained by Range-Doppler (R-D) algorithm. As for its coordinates, they are obtained by CFAR algorithm on the R-D image and calculating the Angle of Arrival(AoA).
Please note that we omit the specific technical details, but readers can learn more in \cite{17}.

\subsection{Point Cloud Model of Gesture}

In this module, we build a new 3D point cloud model of gestures with tendency in space, in order to remove as many interference points as possible.
In Equ.\ref{pi}, we show the set of reflection point information. The $ X, Y, Z, V, I $ respectively represent five features, and the lower corner marker $i$ represents the ith point in the set. 
\begin{equation}
P_{i}=\{X_{i}, Y_{i}, Z_{i}, V_{i}, I_{i}\} \label{pi}
\end{equation}

After collecting the reflective point information data, we are ready to process the data.
We take 30 frames(3 seconds) of initial point data to represent a gesture. By analyzing the trend of the point cloud, we can determine the changing pattern of gestures. Then, we prepare to deal with the initial data. Firstly, we will find the mean of five features of all reflection points of a gesture, taking the average value as a standard point as shown in Equ.\ref{p0}, and then subtract the mean from each reflection point in turn to get D-value point as shown in Equ.\ref{pd}. 
\begin{equation}
P_{0}=\{\bar{X}, \bar{Y}, \bar{Z}, \bar{V}, \bar{I}\} \label{p0}
\end{equation}
\begin{equation}
\delta P_{i}=\{(X_{i}-\bar{X}), (Y_{i}-\bar{Y}), (Z_{i}-\bar{Z}), (V_{i}-\bar{V}), (I_{i}-\bar{I})\} \label{pd}
\end{equation}

Then, we can get a 3D point cloud(pc) model with 5*30*65 size matrix as shown in Equ.\ref{pp}. 
\begin{equation}
\begin{aligned}
PC&=\left( \{F_{1}, F_{2}, F_{3}, F_{3},...F_{j}\}, j=1,2,3...30, \right.\\
&\left. F_{j} = \{P_{0}, \delta P_{1}, \delta P_{2},..., \delta P_{n}\}, n=1,2,3,...64 \label{pp} \right)\\
\end{aligned}
\end{equation}

\begin{figure}
	\centering
	\subfigure[]{
		\includegraphics[width=0.2\textwidth]{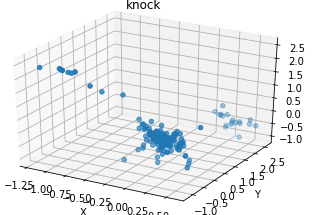}
		\label{pointcloud1}}
	\subfigure[]{
		\includegraphics[width=0.2\textwidth]{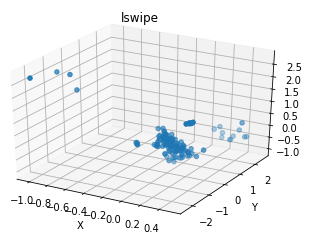}
		\label{pointcloud2}}
	\subfigure[]{
		\includegraphics[width=0.2\textwidth]{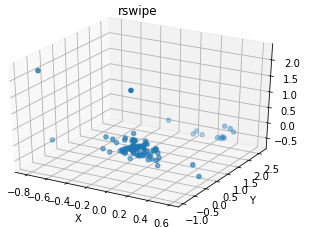}
		\label{pointcloud3}}
	\subfigure[]{
		\includegraphics[width=0.2\textwidth]{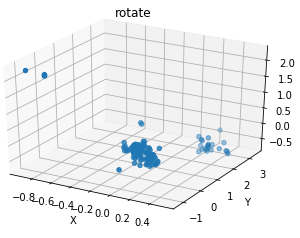}
		\label{pointcloud4}}
	\caption{The point cloud of four gestures: (a) knock; (b) left swipe; (c) right swipe  (d)rotate .}
	\label{pointcloud}
\end{figure}

 As a result, we acquire the new 3D point model, which can focus points in the space near the standard point(shown in Fig.\ref{pointcloud})and input the model to the next CNN.




\subsection{The model of convolutional neural network}

\begin{figure}
	\centering
	\includegraphics[height=2in]{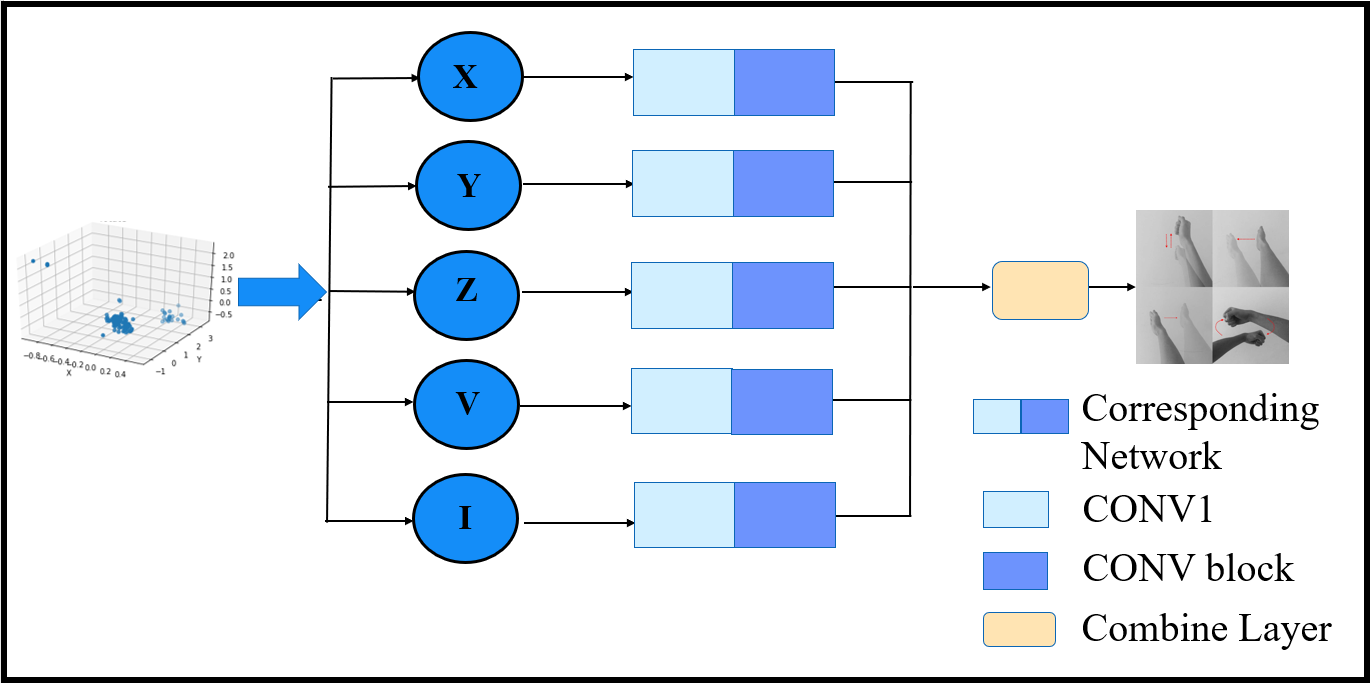}
	\caption{The structure of Neural Network} \label{NN}
\end{figure}

We built a customized multi-branch CNN to process feature information and recognize gestures as shown in Fig.\ref{NN}. 
Since each point in the point cloud contains five features in the information, these features can calculate the valid information for each reflection point we need. Therefore, the CNN is divided into five layers in according to the features, that is, N=$\{X,Y,Z,V,I\}$. Each layer corresponds to $X,Y,Z$ coordinates, velocity and intensity of energy.
Firstly, we establish a ResNet class for each corresponding network. Through the ResNet layer, features can enter their corresponding networks. After passing through the networks, we extract the spatiotemporal information of a feature, but this cannot fully represent the gesture information.
Therefore, the results after each network calculation go into the combine layer, in which these results are combined into a final result, as follows:
\begin{equation}
scr = N(Cat(X(x),Y(y),Z(z),V(v),I(i)))
\end{equation}
Thus, the final result of $jth$ gesture recognized as $j = max(src)$.

\textbf{Implementation and Training:}
The corresponding network contains two parts: CONV1 network and CONV block. CONV1 network has a 7 $\times$ 7 convolution layer with 2 $\times$ 2 strides, and uses batch normalization followed by ReLU activation functions after the layer. It uses the 3 $\times$ 3 max pooling with 2 $\times$ 2 strides. CONV block is a residual block for ResNet with 2 $\times$ 2 strides and 3 $\times$ 3 convolution layers. We connect feature values together into combine layer. The first layer is 3 $\times$ 3 convolution layers with 1 $\times$ 1 strides. We use batch normalization followed by the ReLU activation functions after the layer. We use a layer of fully connected layer with 65,280 input to obtain classification scores. The batch size is 64 and total training epochs are 200. The initial value of learning rate lr is 0.001. For each 200 epoch we set lr = lr $\times$ 0.1. The optimization function of the network is Adam. We implement our network in PyTorch.

\section{Experiment}

\subsection{Dataset}

\begin{figure}
	\centering
	\subfigure[]{
		\includegraphics[height=1in]{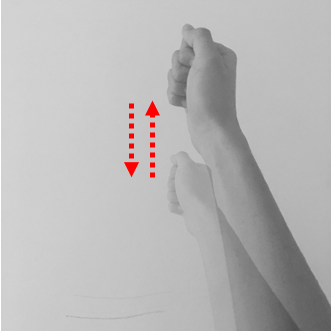}
		\label{Fig.design.click}}
	\subfigure[]{
		\includegraphics[height = 1in]{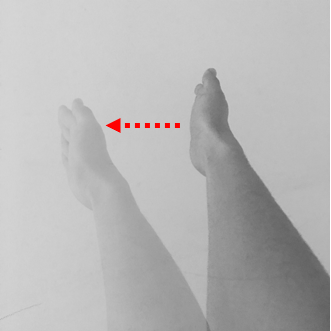}
		\label{Fig.design.lswipe}}
	\subfigure[]{
		\includegraphics[height = 1in]{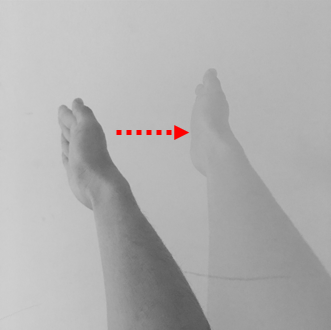}
		\label{Fig.design.rswipe}}
	\subfigure[]{
		\includegraphics[height=1in]{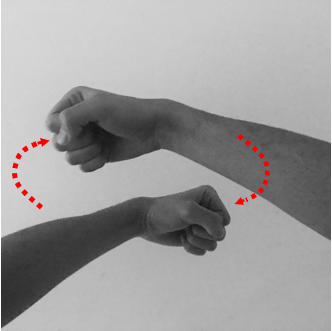}
		\label{Fig.design.clock}}
	\caption{Predefined gestures for interacting with applications: (a) knock; (b) left swipe; (c) right swipe; (d) rotate .}
	\label{Fig.design}
\end{figure}

\textbf{Gesture design:}
In following experiments, participants are invited to perform four gestures in Fig.\ref{Fig.design}: 
knock, left swipe, right swipe and rotate. 
For gesture (a), knock, we ask participants to raise their right arm and tap it up and down twice in the air before returning to their original position.
For gesture (b), left swipe, we ask participants to raise their right arm, sliding it across their chest to the left hand side, and then return to the original position.
For gesture (c), right swipe, we ask participants to raise their right arm to the front of their left arm, sliding it across their chest to the right, and then return to their original position.
For gesture (d), rotate, we ask participants to roll their hands in front of their chest, alternating up and down, for a period of time.
%
%


\textbf{Data collection:}
is performed as follows: Firstly, we place the 
mmWave radar in an empty room that will eliminate other interference. Then we arrange the participants stand 2.4 meters in front of radar and make four different gestures introduced above. In each experiment, we ask 30 different participants, each of whom repeats the same gesture 50 times. 

\subsection{Implementation Details}

	
\textbf{Hardware:}
We utilize a commodity 
mmWave radar that is an integrated single-chip mmWave sensor based on FMCW radar technology operating in 76 to 81GHz bands with continuous linear frequency modulation pulses up to 4GHz. This sensor provides us with rewritable functionality, for changing the internal programming module to get a multi-mode sensor applied for long-range gesture recognition system model. The parameters of the 
radar are shown in Table.\ref{hardware}.

\begin{table}
\centering
\begin{tabular*}{5.9cm}{cc}
		\toprule
		Parameters & Value \\
		\midrule
		Frequency&77GHz  \\
		Antennas &3TX,4RX\\
	    Intermediate Frequency &15MHz\\
	    sampling rate &37.5Msps\\
		Maximum unambiguous range&9.62m\\
		Frame Duration&200msec\\
		TX Power&12dBm\\
		\midrule
\end{tabular*}
	\vspace{0.05in}
	\caption{Radar Parameters }
	\label{hardware}
\end{table}
	
\textbf{Software:}
There are three software components 
to handle 
with data:
Firstly, we use the Code Composer Studio (CCS) 9.1.0 to receive the point cloud data from the mmWave radar. CCS is a 
Integrated Device Electronics (IDE) that supports the embedded microcontroller and processor product line. 
Then, we need Matlab R2017b to save 
the point cloud data as a ``.csv" file. 
At last, we use the Spyder 3.3.6, a simple IDE for python 3.7.3, the convolutional neural network and data processing are both implemented in Python language.

\subsection{Performance analysis}
\label{sec.exp.per}
We design an experiment to test the accuracy under ideal conditions to investigate the model's performance.
Specifically, we perform the standard test on 30 participants(15 men and 15 women). They are asked to make a sequence of four gestures, each repeats 50 times, at a distance of 2.4 metres from the radar.
After all the gesture information is collected, we process the data and get the corresponding off-line accuracy rate. 
By averaging the off-line accuracy of four gestures, we get a total off-line accuracy of 88.11\%.

\textbf{Micro Analysis:}
To make it easier for readers to understand the specific situations of four gestures' accuracies, we built the confusion matrix, shown in Table.\ref{confusion matrix}. 

\begin{table}
	\centering
	\begin{tabular*}{7.5cm}{crrrrrrrrr}
		\toprule
		Types &&knock &&left swipe &&right swipe&&rotate  \\
		\midrule
		knock&&\cellcolor{blue!42.5}85.17\%&&\cellcolor{blue!5.5}11.03\%&&\cellcolor{blue!1.3}2.76\%&&\cellcolor{blue!0.5}1.03\% \\
		left swipe&&\cellcolor{blue!3.5}7.05\%&&\cellcolor{blue!43}86.24\%&&\cellcolor{blue!2.8}5.70\%&&\cellcolor{blue!0.5}1.01\% \\
		right swipe&&\cellcolor{blue!5.5}10.84\%&&\cellcolor{blue!3}5.94\%&&\cellcolor{blue!41}82.17\%&&\cellcolor{blue!0.5}1.05\%\\
		rotate&&\cellcolor{blue!0.35}0.72\%&&70.00\%&&0.00\%&&\cellcolor{blue!49}99.28\%\\
		\midrule
	\end{tabular*}
	\caption{Confusion Matrix of off-line test. In
		the matrix, the probability of the $(i, j)$ element represents that ith
		gesture is recognized as the jth one. The darker the color, the higher the accuracy}
	\label{confusion matrix}
\end{table}

From the confusion matrix, we have several conclusions:
\emph{(i)} The rotate gesture has the highest accuracy rate nearly 99\%, because its unique motion trajectory can be distinguished from other gestures, so it is not easy to be misjudged. This shows that the model has a high accuracy in judging some unique gestures.
\emph{(ii)}The accuracy rates of other three gestures are not high enough, because the transformation modes of the three are similar, especially in a long-range scenario, so they are easy to be mistaked.



\subsection{Practical application analysis}
To verify the performance of the model in real situations, we simulate three real-life scenarios. 
Comparing to experiments in Sec.\ref{sec.exp.per}, we add several chairs and a television to create the effect of the living room in the family. 
We will test the impact caused by these pieces of furniture on our model in following scenes.
 
\begin{figure}
	\centering
	\subfigure[]{
		\includegraphics[width=0.3\textwidth]{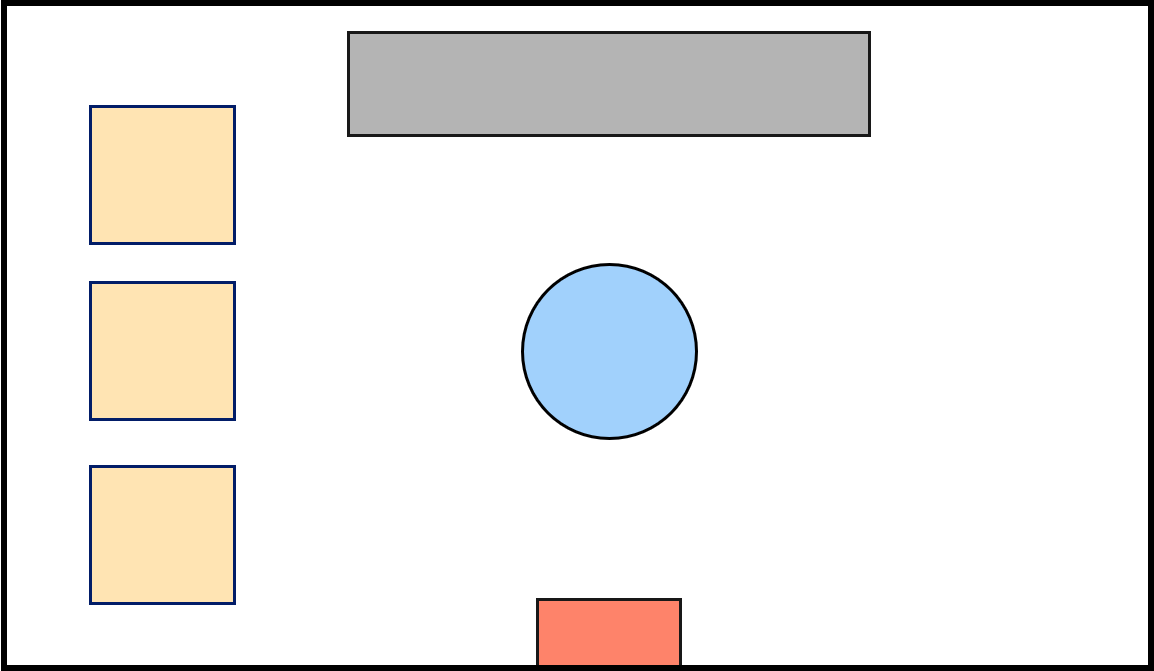}
		\label{Fig.design.click}}
	\subfigure[]{
		\includegraphics[width=0.3\textwidth]{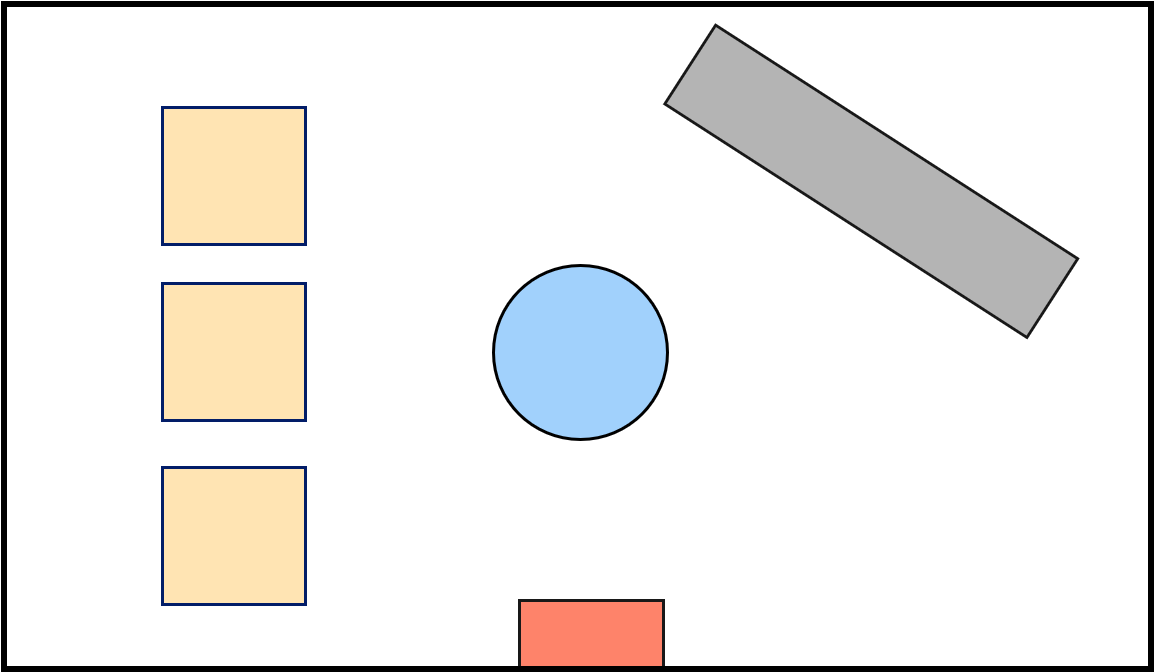}
		\label{Fig.design.lswipe}}
	\subfigure[]{
		\includegraphics[width=0.3\textwidth]{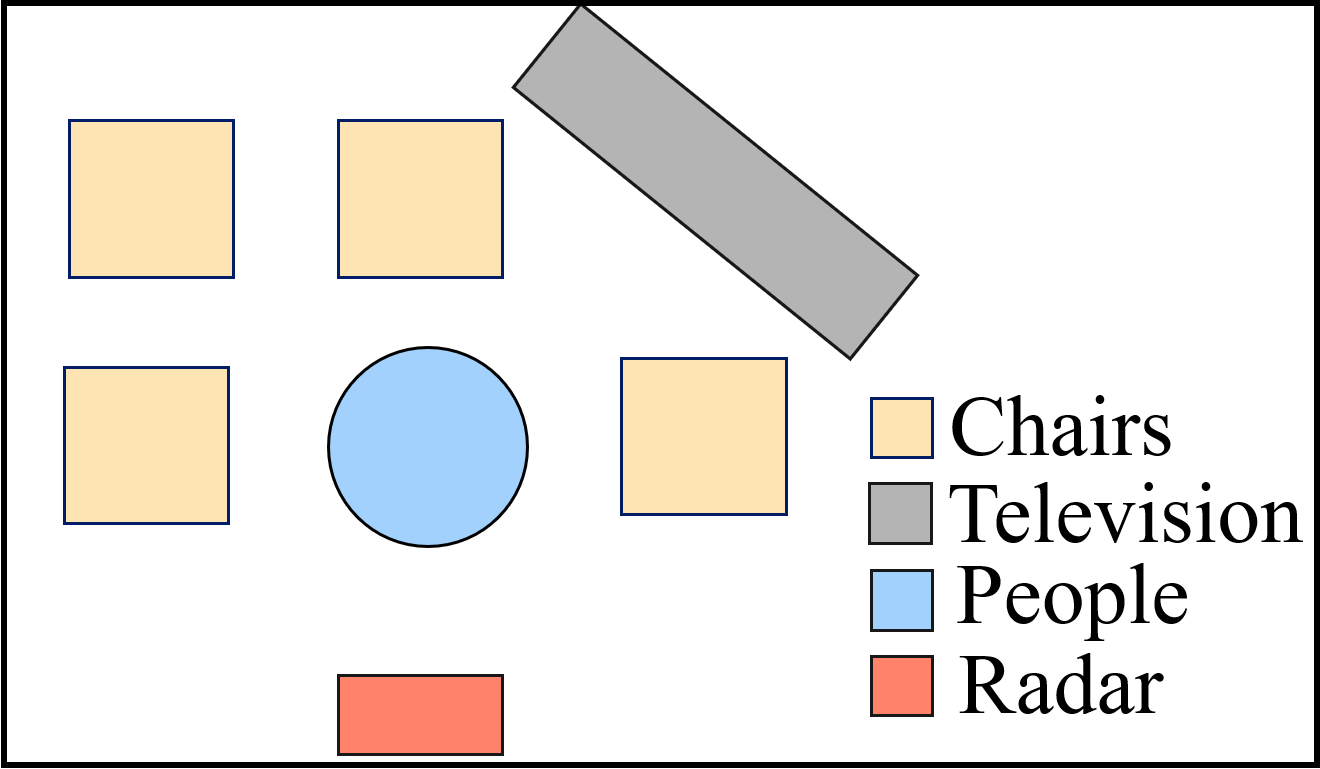}
		\label{Fig.design.rswipe}}
	\caption{Three scenes: (a) scene1; (b) scene2; (c) scene3 .}
	\label{Fig.scene}
\end{figure}

In the first scene, we ask two participants, still standing 2.4 meters from the radar, to repeat each of four gestures 30 times, collecting a total of 60 pieces of gesture data. The room is as shown in Fig.\ref{Fig.scene}.
In the second and third scene, we keep the number of participants and the number and types of gestures still the same. We only change the placement of tables and chairs in the room.

The result is as following:
\begin{table}
	\centering
	\begin{tabular*}{8.15cm}{crrrrrrrr}
		\toprule
		Types/accuracy &knock &&left swipe &&right swipe&&rotate  \\
		\midrule
		scene1&48.33\%&&38.33\%&&51.67\%&&98.00\%\\
		secne2&35.00\%&&5.00\%&&23.33\%&&96.67\%\\
		scene3&38.33\%&&3.33\%&&30.00\%&&98.33\%\\

		\midrule
	\end{tabular*}
	\caption{results of pratical experiment }
	\label{results practical}
\end{table}

From Table.\ref{results practical}, we can find following three insights:
\emph{(i)}Rotate's accuracy has remained roughly the same, suggesting that the uniqueness of the gesture mode gives it a high degree of accuracy. This kind of gestures is less affected by the external environment, and can be applied in real life.
\emph{(ii)}The accuracy of the first three gestures has decreased greatly, which is not enough to support the application of the model in real life. We think that this model is greatly affected by external environment factors, and does not have a good performance for gestures that are not easy to distinguish.
\emph{(iii)}However in these three scenarios, we only collected 60 pieces of data for each gesture, so the lack of data may also has contributed to the sharp decline in accuracy
\section{Discussion}
In this section, we have a case-by-case discussion on factors which might affect the model. The distance factor, the multi-people factor and the real-time situation are discussed in turn. We explore the performance and practical application of our model under the influence of different factors. These discussions allow us to evaluate and test the model more thoroughly.
\subsection{Distance influence}
\begin{figure}
\centering
\includegraphics[width=7cm]{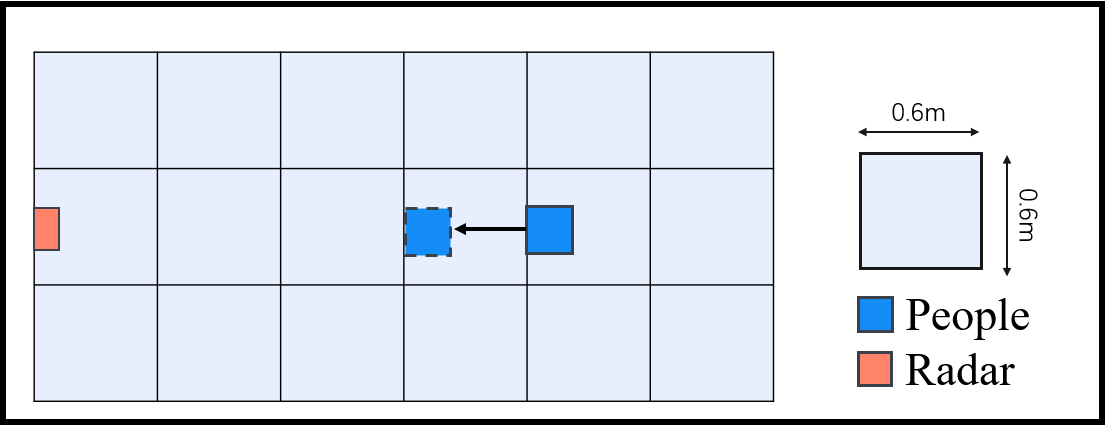}
\caption{The room was divided into bricks. } \label{the roomdiagram}
\end{figure}

During the study, we find that the distance between hand and radar is an important factor that affects the accuracy of model judgment. Here, we divide the room area with bricks, each brick is a square which has the side length of 0.6 meters, as shown in Fig.\ref{the roomdiagram}.
In order to better explore the distance influence on the model, we take the number of bricks as a variable to verify whether the accuracy of the model in different blocks will be affected. First, there are 80 participants in this experiment, each of whom stands at the 5th brick and repeats each of four gestures 50 times, so we collect 400 pieces of data for each gesture.
The accuracy in Sec.\ref{sec.exp.per} is about the same as our previous experiment. 
Therefore, we change the distance between participants and the radar, asking them stand at the 4th brick and keep everything else the same.

We use the model of the 5th brick to test whether the data collected from the 4th brick is accurate. The result is that the data collected from the 4th brick is not accurate.
However, when we input the data collected by participants standing at the 4th brick into the model, and we test participants standing on the 4th brick and the 5th block respectively, and then find that the model could run normally with high accuracy (listed in Table.\ref{fourbricks}).


\begin{table}
	\centering
	\begin{tabular*}{9.2cm}{crrrrrrrrr}
		\toprule
		Scenes &&knock &&left swipe &&right swipe&&rotate  \\
		\midrule
		5th brick(origin model)&&65.00\%&&79.25\%&&87.00\%&&99.25\%\\
		4th brick(origin model)&&3.50\%&&22.25\%&&12.00\%&&98.00\%\\
		5th brick(new model)&&61.75\%&&82.50\%&&88.00\%&&99.00\%\\
		4th brick(new model)&&63.25\%&&80.00\%&&83.33\%&&99.85\%\\
		
		\midrule
	\end{tabular*}
	\caption{The results of distance influence. The origin model only has the data of 5th brick and the new model has the data of both 5th and 4th bricks. }
	\label{fourbricks}
\end{table}

From Table.\ref{fourbricks}, we can find that the model is very sensitive to distance factor, and the gesture data from different distances will lead to different judgment results. Therefore, in practical application, we need to collect as much information as possible from different distances so that the model can accurately judge gestures at different positions. 

In our future work, we manage to reduce the distance sensitivity of the model and realize accurate judgment of gestures at different positions.

\subsection{Multiple people influence}
\begin{figure}
	\centering
	\begin{minipage}{6cm}
		\centering
		\includegraphics[width=\textwidth]{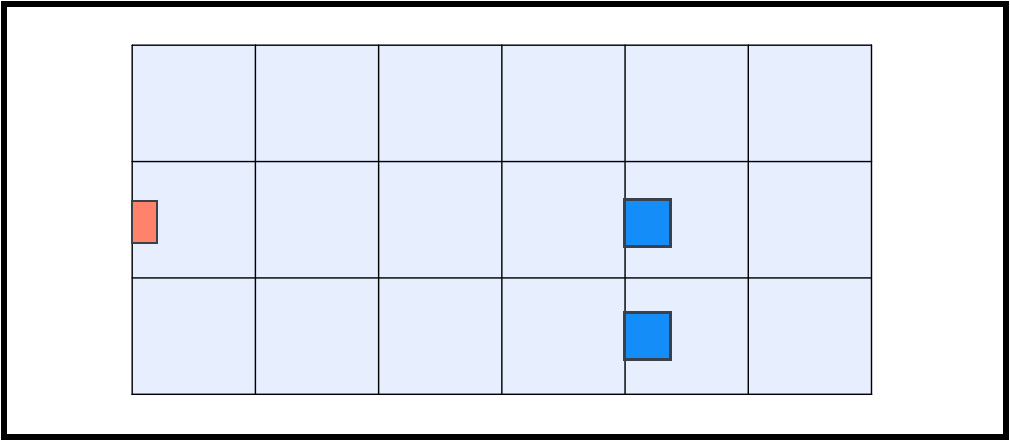}
		\label{Fig.design.click}
		 \centerline{(a) row}
	\end{minipage}
    \begin{minipage}{6cm}
    	\centering
    	\includegraphics[width=\textwidth]{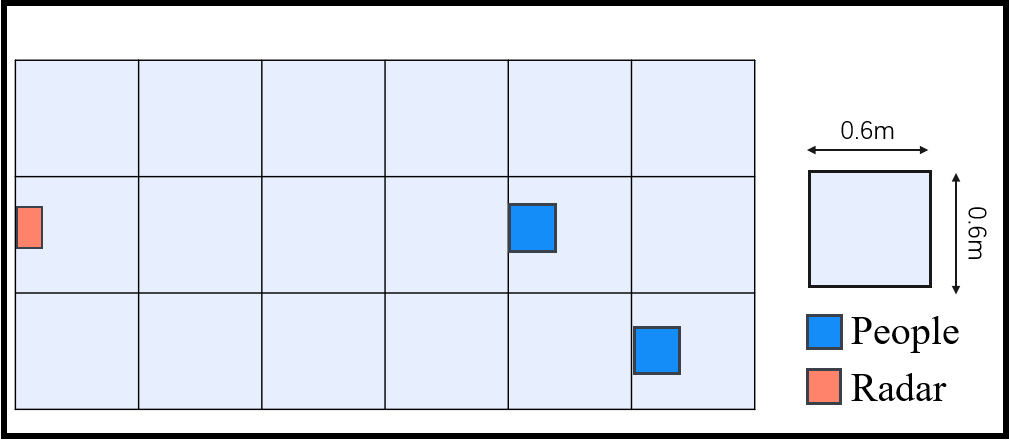}
        \label{Fig.design.click}
        \centerline{(b) front}
    \end{minipage}
	\caption{the two scenarios: (a)two praticipants stand in one row (b)one praticipant stands in front of another} \label{rows}
\end{figure}

We apply the model to the multi-people environment, and design two scenarios shown as Fig.\ref{rows}.

In this experiment, we recruit three participants and divide each two of them into three groups. Each group will experiment with two scenes separately. Among them, the participant who stands on the 5th brick is set as the target person we want to identify, and the other participant is the interferer. Each group asked to do each of the four gestures 30 times simultaneously, so we can get 90 pieces of data of each gesture.
The experiment results of two scenarios are shown in Table.\ref{multi}.

\begin{table}
	\centering
	\begin{tabular*}{8.7cm}{crrrrrrrr}
		\toprule
		Scenes &&knock &&left swipe &&right swipe&&rotate  \\
		\midrule
		Multi-people(row)&&37.78\%&&2.22\%&&22.22\%&&100.00\%\\
		Multi-people(front)&&47.78\%&&1.00\%&&17.78\%&&98.89\%\\	
		\midrule
	\end{tabular*}
	\caption{results of multiple people }
	\label{multi}
\end{table}

From the results, we can find that only rotate can the model judged correctly with a high accuracy.
However, the accuracy of another three gestures is too low to be used.
Thus, this model hasn't accomplished to recognize multiple people's concurrent gestures.
It is also what we will devote to in the future.
\subsection{Real-time consideration}
In the real situation, the model usually encounters a lot of interferences, which leads to decreased accuracy in judging gestures. 
Therefore, we consider several interference actions, 
\ie, walk, run, and some tiny gestures which are similar to our model gestures.
In this experiment, we ask 80 participants to do each of these two interference actions (distance and other requirements remained the same), and to repeat each action 50 times. We can get 400 pieces of data for each action. 

In consequence of not considering the model of interference before this experiment, the model will mistake these interference actions into different gestures. The results are shown as Table.\ref{interference} :

\begin{table}
	\centering
	\begin{tabular*}{7.4cm}{crrrrrrr}
		\toprule
		Scenes &knock &&left swipe &&right swipe& &rotate  \\
		\midrule
		Tiny&4.75\%&&24.25\%&&5.25\%& &65.75\%\\
		Walk+Run&25.25\%&&0.75\%&&1.75\%& &72.25\%\\	
		\midrule
	\end{tabular*}
	\caption{model without interference actions }
	\label{interference}
\end{table}

From the Table.\ref{interference}, some tiny gestures may be mistaken into rotate or other gestures.
The action of walk or run is also an interference for our model. This will cause the accuracy of the model decreased sharply.
We specifically collect 1,600 pieces of data on these two types of interference and feed them into the model. Later, when we test the gesture with the interference actions again, the results from the Table.\ref{without interference} show that the model is able to automatically identify and removes the interference, then identifying the correct gesture.
\begin{table}
	\centering
	\begin{tabular*}{7.4cm}{crrrrrrr}
		\toprule
		Scenes &knock &&left swipe &&right swipe& &rotate  \\
		\midrule
		Tiny&80.75\%&&75.00\%&&78.20\%& &97.75\%\\
		Walk+Run&84.25\%&&85.25\%&&80.25\%& &98.25\%\\	
		\midrule
	\end{tabular*}
	\caption{model with interference actions }
	\label{without interference}
\end{table}
\section{Conclusion}
In this paper, we introduce a long-range gesture recognition model based on mmWave sensing. We utilize a mmWave radar to collect spatial and temporal characteristics of gestures, and then input gesture data into a customized convolutional neural network for automatic gesture recognition.
On this basis, we also carry out several real scenario experiments to verify 
the practicability of the model. 
With the rapid development of 5G technology, mmWave frequency band will be 
fully developed and utilized 
in the future.
Therefore, we believe mmWave sensing technology will provide a wider range of services for mankind.

\end{document}